\documentclass{aa}  

\usepackage{graphicx}
\usepackage{natbib}
\usepackage{txfonts}
\usepackage[colorlinks=true,citecolor=blue]{hyperref}%

\usepackage{xcolor}
\newcommand{\for}{$f(R)$}
\newcommand{\Alens}{A_\textrm{lens}}

\newcommand{\logfr}{\log\lvert f_{R_0} \lvert}
\newcommand{\Planck}{{\em Planck}}
\newcommand{\lcdm}{$\Lambda$CDM}
\newcommand{\react}{{\it ReACT}}
\newcommand{\mgclass}{{\it MGCLASS}}
\newcommand{\nside}{N_\textrm{SIDE}}
\newcommand{\namaster}{{\sl NaMaster}}
\newcommand{\halofit}{{\sl halofit}}
\newcommand{\emcee}{{\sl emcee}}
\newcommand{\twotimestwo}{$2\times 2$\,pt}
\newcommand{\threetimestwo}{$3\times 2$\,pt}

\begin{document}

   \title{Constraining f(R) gravity with cross-correlation of galaxies and cosmic microwave background lensing}

    \titlerunning{Constraining f(R) gravity with cross-correlation of galaxies and CMB lensing}

   \author{Raphaël Kou\inst{1}, Calum Murray\inst{1}\and James G. Bartlett\inst{1}}

   \authorrunning{R. Kou, C. Murray, J. G. Bartlett}

    \institute{Université Paris Cité, CNRS, Astroparticule et Cosmologie, F-75013 Paris, France} 

    \date{Received / Accepted}

 
  \abstract{We look for signatures of the Hu-Sawicki \for\ modified gravity theory, proposed to explain the observed accelerated expansion of the universe; in observations of the galaxy distribution, the cosmic microwave background (CMB), and gravitational lensing of the CMB. We study constraints obtained by using observations of only the CMB primary anisotropies, before adding the galaxy power spectrum and its cross-correlation with CMB lensing. We show that cross-correlation of the galaxy distribution with lensing measurements is crucial to breaking parameter degeneracies, placing tighter constraints on the model. In particular, we set a strong upper limit on $\logfr<-4.61$ at $95\%$ confidence level. This means that while the model may  explain the accelerated expansion, its impact on large-scale structure closely resembles General Relativity. Studies of this kind with future data sets will probe smaller potential deviations from General Relativity.} 


   \maketitle

\section{Introduction}

The cause of the late-time accelerated expansion of the Universe is one of the most profound problems facing modern cosmology ~\citep{Riess98,Perlmutter99}. Many theories have been proposed to explain this phenomena, the most popular due to its apparent simplicity is $\Lambda$, the cosmological constant. However, the interpretation of the cosmological constant as the energy of the vacuum results in theoretical predictions that are at least 55 orders of magnitude too large~\citep[e.g.,][]{Carroll04,Sola13}, motivating the study of alternative explanations. One possibility is new gravitational physics~\citep{Carroll04}. In addition to causing the accelerated expansion,  modifications to gravity may alter structure formation in the Universe. We constrain a particular model of \for\ gravity, which can explain the accelerated expansion of the Universe, using observations of galaxy clustering, weak gravitational lensing of the cosmic microwave background (CMB), and temperature and polarization information from the CMB.

Deviations from General Relativity (GR) are tightly constrained on the scales of our Solar System~\citep{Everitt11,Will14}. Therefore, models of modified gravity must at the same time satisfy these constraints on small-scales, whilst modifying gravity on large scales to explain the cosmic acceleration. ~\citet{Carroll04} presented a general class of models that can drive cosmic acceleration, by replacing the linear dependence of the Einstein-Hilbert action on the Ricci scalar $R$ with a non-linear function of $R$ ($R \rightarrow R + f(R)$). ~\citet{Hu-Sawicki} (in the following, HS) presented a class of \for\ models capable of explaining the cosmic acceleration and evading the strong Solar System constraints through a chameleon mechanism~\citep{Khoury04,Navarro07,Faulkner07}.

Constraints have been placed on HS \for\ gravity using many different complementary observations. Such observations constrain $f_{R_0}$, the value of the cosmological field today, that we introduce in more detail in Sect.~\ref{sec:for}. In particular, on cosmological scales HS \for\ has been constrained by~\citet{Cataneo15}, who obtained the constraint $\logfr<-4.79$ at the $95\%$ confidence level, using cluster number counts in addition to CMB, supernovae and BAO data. \citet{Hu16} also found $\logfr<-4.5$ using the CMB (temperature, polarisation and lensing), supernovae, BAO and galaxy weak lensing measurements. \citet{Hojjati16} obtained the upper bound $\logfr<-4.15$ at the $95\%$ confidence level using similar observations. 

The strongest constraints come from galactic scales. \citet{Naik19} were able to exclude $\logfr>-6.1$ using galaxy rotation curves, and~\citet{Desmond20} excluded $\logfr>-7.85$ based on the analysis of galaxy morphology. Astrophysical and cosmological constraints on HS \for\ gravity can also be found in the review by~\citet{Lombriser14}. Finally,~\citet{Casas23} forecasts the constraints that will be achievable using observations from {\em Euclid}. Despite the strong constraints on HS \for\ gravity from galactic studies, it is still a useful model to explore deviations from GR on cosmological scales.

Many different tools have been developed to predict the matter power spectrum in HS \for\ gravity. Boltzmann codes that calculate the linear matter power spectrum are {\sl mgcamb}~\citep{mgcamb1,mgcamb2,mgcamb3,mgcamb4} and \mgclass~\citep{mgclass}. There are several simulation based emulators of the matter-power spectrum into the mildy non-linear regime ~\citep{Winther19,Ramachandra21,Arnold22,emantis}, and \react~\citep{Bose:2020wch,Bose:2022vwi}, which uses a halo model reaction framework validated on N-body simulations.

In the following section we review HS \for\ gravity. In Sect.~\ref{sec:data} we introduce the observations used in our analysis, and then in Sect.~\ref{sec:method} we overview our methodology; the estimation of the angular power spectrum, our covariance matrix estimation, and our likelihood. Results are presented in Sect.~\ref{sec:results}, before the conclusions in Sect.~\ref{sec:conclude}.

\section{Hu-Sawicki f(R) gravity}\label{sec:for}

In \for\ theories of gravity, the Einstein-Hilbert action is modified such that $R \rightarrow R + f(R)$; therefore, the action becomes,
\begin{align}
    S = \int d^4x\sqrt{-g}\left[\frac{R+f(R)}{2\kappa^2}+\mathcal{L}_m\right],
\end{align}
where $R$ is the Ricci scalar, $\kappa=8\pi G$ with $G$ the gravitational constant (and the speed of light set to $1$), $g$ is the determinant of the spacetime metric, $\mathcal{L}_m$ is the matter Lagrangian, and $f$ is a function of the Ricci scalar. In HS~\citep{Hu-Sawicki}, $f(R)$ follows a broken power law,
\begin{align}\label{eq:f_of_R}
    f(R) = -m^2\frac{c_1(R/m^2)^n}{c_2(R/m^2)^n+1},
\end{align}
where $m$ is a mass scale given by $m^2=\kappa^2\bar{\rho}_m/3$ with $\bar{\rho}_m$ the mean matter density of the Universe, and $c_1$, $c_2$ and $n$ are three dimensionless constants. The derivative of $f$ with respect to the Ricci scalar $R$ is denoted,
\begin{align}\label{eq:f_R}
    f_R = \frac{df(R)}{dR}=-\frac{nc_1\left(\frac{R}{m^2}\right)^{n-1}}{\left(c_2\left(\frac{R}{m^2}\right)^n+1\right)^2},
\end{align}
and can be interpreted as a new scalar field. \citet{Hu-Sawicki} showed that a background close to \lcdm\ can be recovered by imposing, 
\begin{align}\label{eq:c1_c2}
    \frac{c_1}{c_2} = 6\frac{\Omega_\Lambda}{\Omega_m},
\end{align}
where $\Omega_\Lambda$ and $\Omega_m$ are the present-day dark energy and matter densities (divided by the critical density) in the \lcdm\ cosmology. Imposing this relation, there remain only two free parameters in Eq.~(\ref{eq:f_of_R}): $n$ and either $c_1$ or $c_2$. In the high curvature regime ($R \gg m^2$), which~\citet{Oyaizu08} showed to be the appropriate regime, Eq.~(\ref{eq:f_R}) can be written as,
\begin{align}
    f_R = -n\frac{c_1}{c_2^2}\left(\frac{m^2}{R}\right)^{n+1},
\end{align}
which, evaluated at the present-day background, leads to,
\begin{align}
    \frac{c_1}{c_2^2} = -\frac{1}{n}f_{R_0}\left(\frac{R_0}{m^2}\right)^{n+1}.
\end{align}
The $f_{R_0}$ parameter denotes the background value of $f_R$ at the present time, which we choose as our free parameter to constrain the model of HS \for. Additionally, we fix $n=1$.

For the small values of $f_{R_0}$ probed in this work, the background expansion of the Universe is indistinguishable between \for\ and that of a cosmological constant. Instead, we constrain \for\ through its impact on the growth of structure. This can be seen by looking at the modified Poisson equation in \for,
\begin{equation}
\nabla^2 \Phi = \frac{\kappa}{2} a^2 \delta \rho_m - \frac{1}{2} \nabla^2 f_R,
\end{equation}
where $a$ is the cosmological scale factor and $\delta \rho_m \equiv \rho_m - \bar{\rho}_m $. We see directly that $f_R/2$ can be seen as the potential of the modified gravity force. As mentioned in the introduction, this modified Poisson equation approaches the GR expression within the Solar System through the chameleon mechanism~\citep{Khoury04,Hu-Sawicki}.

It is worth noting that unlike other theories of modified gravity, HS \for\, has little effect on the propagation of light in the weak-field limit~\citep[for example][]{Hojjati16}.

\section{Data}\label{sec:data}

\subsection{BOSS galaxies}

\begin{figure}
  \resizebox{\hsize}{!}{\includegraphics{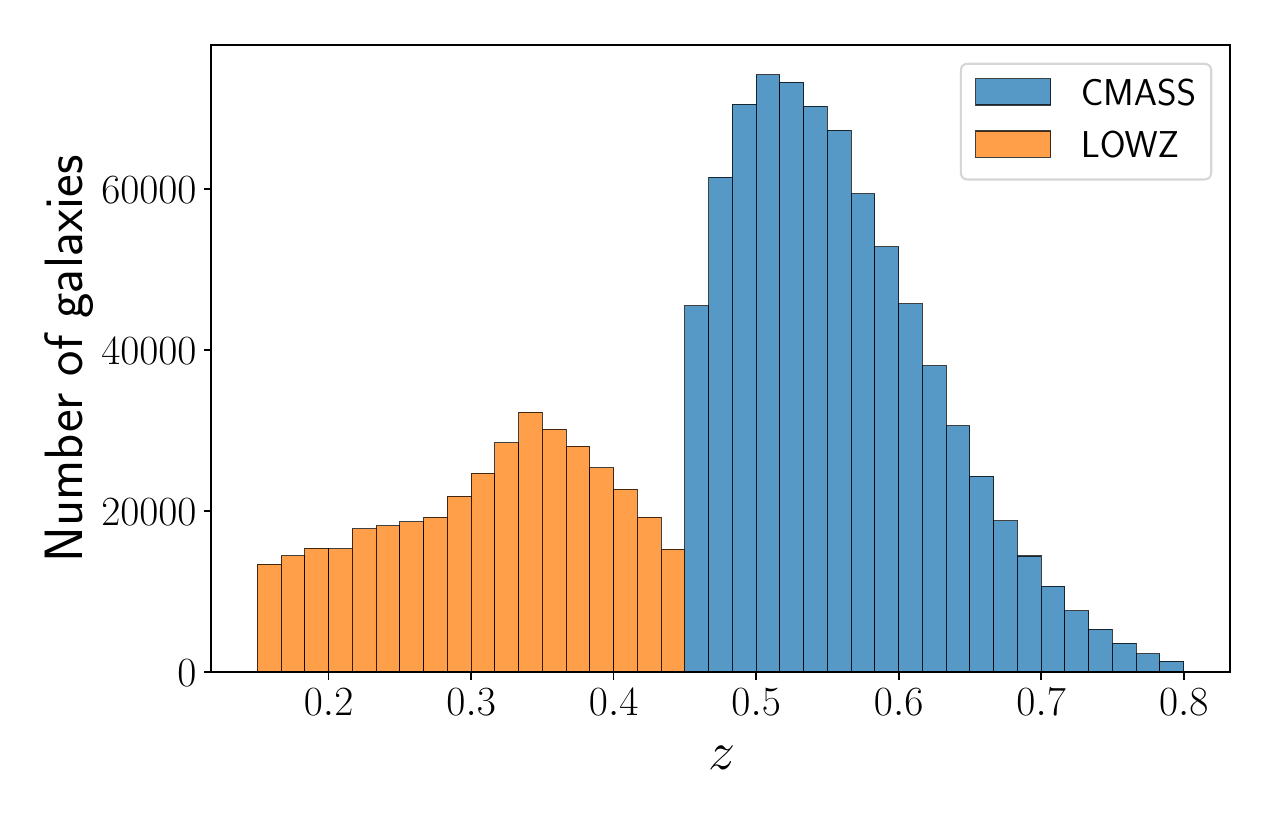}}
  \caption{CMASS (blue) and LOWZ (orange) redshift distributions.}\label{fig:galaxy_distribution}
\end{figure}

We use the DR12 data release of the BOSS survey from the SDSS collaboration \citep{Alam15}. This large-scale spectroscopic survey was divided into two subsamples, LOWZ and CMASS. LOWZ contains galaxies at low redshift, up to approximately $z\simeq 0.45$, while CMASS contains higher redshift galaxies, (roughly up to $z\simeq 0.8$) and was constructed to create a sample of galaxies with approximately constant stellar mass. As in~\citet{Loureiro19}, we restrict these samples to $0.15<z<0.45$ and $0.45<z<0.8$ for LOWZ and CMASS, respectively, such that the two samples do not overlap in redshift. This choice allows us to neglect the covariance between galaxies of LOWZ and CMASS. Using these redshift ranges, and after masking regions of low completeness, the two samples contain $366,576$ and $751,067$ galaxies. The redshift distribution of the two samples is shown in Fig.~\ref{fig:galaxy_distribution}.

The mask and map making is identical to that within ~\citet{Kou22}, which follows ~\citet{Reid16} and~\citet{Loureiro19}. We transform the {\sl MANGLE}\footnote{https://space.mit.edu/$\sim$molly/mangle/} \citep{Swanson08} acceptance and veto masks, provided with the galaxy catalogs, into high resolution binary masks in {\sl HEALPix}\footnote{http://healpix.sf.net} \citep{healpix, Zonca2019} format with $\textrm{N}_\textrm{SIDE}=8192$. The acceptance mask represents the completeness of the observations, while the veto mask excludes regions that could not be observed. A first cut is made to exclude regions with completeness below $0.7$, before degrading the resolution of the mask to $\textrm{N}_\textrm{SIDE}=4096$. A second cut is then applied, such that pixels with completeness below $0.8$ are rejected. Finally, the galaxy maps are computed by summing the weighted number of galaxies in each pixel, divided by the completeness of the pixel. The weight $w_\textrm{tot}$ that is applied to each galaxy takes into account a number of observational effects, including fibre collisions, redshift failures, stellar density and seeing conditions \citep[more details can be found in][]{Ross12}. The galaxy overdensity map is,
\begin{align}\label{eq:overdensity_equation}
    \delta_p = \left(\frac{n_p}{\bar{n}}-1\right),
\end{align}
where,
\begin{align}
    n_p = \left\{
    \begin{array}{ll}
        \frac{1}{C_\textrm{pix}^p}\sum_{i\in p} w_\textrm{tot}^i & \mbox{if } C_\textrm{pix}^p > 0.8 \\
        0 & \mbox{otherwise},
    \end{array}
\right.
\end{align}
where $C_\textrm{pix}^p$ is the completeness in pixel $p$.

\subsection{CMB temperature and polarization observations}\label{sec:cmb}

We use observations of the CMB temperature and polarization anisotropies from the \Planck\ satellite, which observed the CMB for about $29$ months and covered the full sky. In this work, we directly make use of the likelihood code provided~\citep{Planck_likelihood} and whose cosmological results were analyzed in~\citet{Planck_cosmo18}.

\subsection{CMB lensing convergence map} \label{sec:lensing}

The observed CMB fluctuations are distorted as the CMB photons traverse the Universe because of gravitational lensing. An observational consequence of this is the correlation between different multipoles in both the temperature and polarization anisotropies, which would not be present in the unlensed CMB. The CMB lensing potential can therefore be reconstructed from such correlations (see \cite{Lewis06} for a comprehensive review).

We use the CMB lensing convergence map released by~\citet{Aghanim20}. This map was obtained using a minimum variance quadratic estimator based on temperature and polarization maps. This map covers about $67\%$ of the sky and led to the detection of lensing at $40\sigma$. The map is provided with resolution $\nside=4096$, together with the associated mask with $\nside=2048$. 

\section{Methodology}\label{sec:method}

\subsection{Theoretical angular power spectra}\label{sec:effect_for_and_theory}

We calculate the matter power spectrum in \for\ using two different codes, \mgclass\ and \react. 

\mgclass~\citep{mgclass} is a modified version of the Boltzmann code {\sl CLASS}~\citep{CLASS} in which the equations of the linear perturbation theory are changed to take into account  modifications to gravity. It can therefore be used to predict the linear matter power spectrum. 

\react~\citep{Bose:2020wch,Bose:2022vwi} gives predictions for the non-linear matter power spectrum in beyond \lcdm\ cosmologies, including $w\textrm{CDM}$, \for\ and DGP gravity. \react\ uses a halo model based approach described in~\citet{Cataneo:2018cic} such that,
\begin{align}\label{eq:react}
    P_\textrm{NL}(k,z) = \mathcal{R}(k,z)P_\textrm{NL}^\textrm{pseudo}(k,z),
\end{align}
where $P_\textrm{NL}$ is the non-linear matter power spectrum in modified gravity, and $P_\textrm{NL}^\textrm{pseudo}$ is the so-called non-linear ‘‘pseudo-power spectrum''. This pseudo-power spectrum is defined as a \lcdm\ power spectrum with initial conditions chosen such that the \lcdm\ linear matter power spectrum matches the modified gravity linear matter power spectrum at a given redshift. This choice was made in order to ensure that the halo mass function in \lcdm\ and in the modified gravity theory are similar (which is anticipated since they have been defined to have exactly the same linear matter power spectrum). 

The remaining term in Eq.~(\ref{eq:react}), $\mathcal{R}$ is called the reaction and describes how the \lcdm\ matter power spectrum changes due to the modifications to gravity. The reaction is calculated using the halo model and 1-loop perturbation theory. More details can be found in~\citet{Cataneo:2018cic} and~\citet{Bose:2022vwi}. When using \react\ to predict the non-linear modified gravity matter power spectrum, it is required to provide a reliable non-linear \lcdm\ matter power spectrum, for which we use the halo model based {\sl HMCode}~\citep{Mead15}. 

\begin{figure}
  \resizebox{\hsize}{!}{\includegraphics{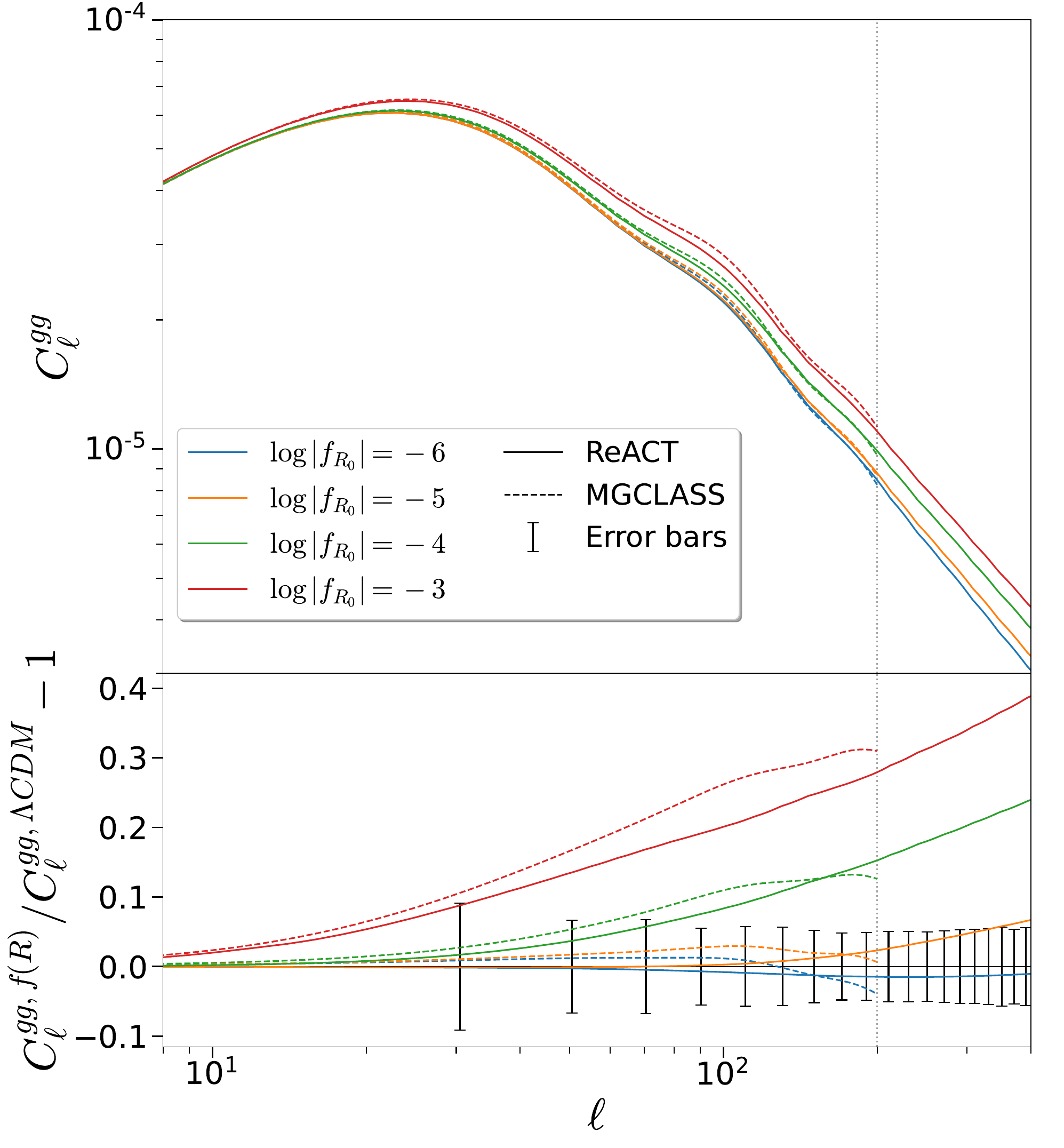}}
  \caption{Effect of changing the value of $\logfr$ on the auto-power spectrum of CMASS, with all other parameters fixed. The bottom panel shows the difference relative to the non-linear power spectrum in \lcdm, together with the $1\sigma$ uncertainties of CMASS. Plain lines are obtained using \react\ and the dashed lines with \mgclass. The grey dotted line shows the limit between the linear and non-linear regimes. We limit our analysis to the linear regime, but show the theoretical predictions with \react. We do not show the predictions with \mgclass, which fail in this regime.}\label{fig:cl_gg_variations_fr}
\end{figure}

For our observations we compute the galaxy auto power spectrum $C_\ell^{gg}$, the CMB lensing convergence auto power spectrum $C_\ell^{\kappa\kappa}$ and the cross-correlation between the two $C_\ell^{\kappa g}$. We also compute the CMB temperature and polarization power spectra, which are sensitive to \for\ gravity through the integrated Sachs-Wolfe effect (ISW) and gravitational lensing. The CMB temperature, polarization and convergence power spectra are  predicted by \mgclass. 

For the galaxy auto and cross power spectra, we use the matter power spectrum prediction from either \react\ or \mgclass. The angular power spectra are then modeled using the Limber approximation ~\citep{Limber53},
\begin{align}
    C_\ell^{gg} &= \int \frac{dz}{c}\frac{H(z)}{\chi^2(z)}W_g^2(z)P_m \left(k=\frac{\ell}{\chi(z)},z \right) \\
    C_\ell^{\kappa g} &= \int \frac{dz}{c}\frac{H(z)}{\chi^2(z)}W_g(z)W_\kappa(z)P_m \left(k=\frac{\ell}{\chi(z)},z \right),
\end{align}
where $H(z)$ is the Hubble parameter at redshift $z$, $c$ is the speed of light, $\chi$ is the comoving distance, $P_m$ is the matter power spectrum, and $k$ is the comoving wavenumber. Finally, $W_g$ and $W_\kappa$ are the galaxy and CMB lensing kernels,
\begin{align}\label{eq:cl_gg}
    W_g(z) &= \frac{b_g}{n_\textrm{tot}}\frac{dn}{dz}\\
    \label{eq:cl_kg}
    W_\kappa(z) &= \frac{3}{2}\Omega_mH_0^2\frac{(1+z)}{H(z)}\frac{\chi(z)}{c}\left(\frac{\chi(z_*)-\chi(z)}{\chi(z_*)}\right).
\end{align}
Here, $b_g$ is the galaxy bias, $(1/n_\textrm{tot}) ( dn/dz )$ is the normalized galaxy redshift distribution, $H_0$ is the present value of $H$, $\Omega_m$ is the matter density parameter, and $z_*$ is the redshift of the surface of last scattering. 

In Fig.~\ref{fig:cl_gg_variations_fr} we show the effect of changing $\logfr$ on the galaxy angular power spectrum, using the galaxy redshift distribution of CMASS (see Sect.~\ref{sec:data} for more details). It can be seen that HS \for\ gravity increases the formation of structure on small scales, leading to more power in the angular power spectrum at larger multipoles. For a given value of $\logfr$, \mgclass\ predicts slightly more power than \react, except at the highest multipoles, where power might be missing in the prediction of \mgclass\ as \mgclass\ only predicts the linear matter power spectrum. This is also the reason why we do not show the predictions using \mgclass\ after the dotted grey line marking the transition into the non-linear regime. The non-linear regime is not used in our analysis as we do not use a theoretical model that can reliably model the galaxy bias in this regime.

\subsection{Angular power spectra estimation}

The angular cross-correlation power spectrum of two fields $A$ and $B$ is defined as,
\begin{align}
    \langle a_{\ell m} b^*_{\ell' m'}\rangle = \delta_{\ell \ell'}\delta_{m m'}C_\ell^{AB},
\end{align}
where $a_{\ell m}$ and $b_{\ell' m'}$ are respectively the spherical harmonic coefficients of fields $A$ and $B$, $b^*$ denotes the complex conjugate of $b$, $\delta$ is the Kronecker symbol, and $\langle\,\rangle$ is the ensemble average. Given full sky coverage, this power spectrum can be estimated using,
\begin{align}\label{eq:Cl_full_sky}
    \hat{C}_\ell^{AB} = \frac{1}{2\ell+1}\sum_{m=-\ell}^\ell a_{\ell m}b_{\ell m}^*.
\end{align}
In practice, however, fields $A$ and $B$ are not observed on the full sky, but on a limited sky fraction defined by their masks $\mathcal{W}^A$ and $\mathcal{W}^B$, such that what we really observe is $\tilde{A}(\mathbf{\hat{n}})=\mathcal{W}^A(\mathbf{\hat{n}})A(\mathbf{\hat{n}})$ and $\tilde{B}(\mathbf{\hat{n}})=\mathcal{W}^B(\mathbf{\hat{n}})B(\mathbf{\hat{n}})$. It can then be shown~\citep{Hivon02,Brown05} that taking the ensemble average of Eq.~(\ref{eq:Cl_full_sky}) with the observed fields $\tilde{A}$ and $\tilde{B}$ gives,
\begin{align}\label{eq:pseudo_cl}
    \langle \hat{C}_\ell^{\tilde{A}\tilde{B}} \rangle = \sum_{\ell'}M_{\ell \ell'}C_\ell^{AB},
\end{align}
where $M_{\ell \ell'}$ is a coupling matrix that depends on the masks $\mathcal{W}^A$ and $\mathcal{W}^B$. It is then possible to recover an unbiased estimate of $C_\ell^{AB}$ by inverting the coupling matrix. 

In our analysis we use the public code \namaster\footnote{https://github.com/LSSTDESC/NaMaster}\citep{Alonso19} to estimate the power spectra of CMASS and LOWZ, as well as their cross-correlation with the CMB lensing convergence map from \Planck. For the CMB lensing auto-correlation, we take directly the \Planck\ lensing likelihood. We also apodize the CMB lensing mask with a scale of $10$\,arcmin, and we have verified that the estimated power spectra do not depend on the apodization scale.

\subsection{Noise removal}

The measured galaxy angular power spectrum is biased by the shot-noise contribution. Therefore, this contribution is subtracted from the estimated power spectrum. The galaxy shot noise is given by,
\begin{align}
    N_\ell^{gg} = \frac{4\pi f_\textrm{sky}}{N},
\end{align}
where $N$ is the weighted number of galaxies in each sample. 

\subsection{Scale cuts}\label{sec:scales_and_binning}

As mentioned in Sect.~\ref{sec:effect_for_and_theory}, for our cosmological constraints we only use power spectra in the linear regime. This limitation mainly comes from the fact that we use a linear galaxy bias and that this simple modeling is not reliable in the non-linear regime. To determine the maximum multipole that can be used, we follow the approach of~\citet{Loureiro19}. Namely, we use the fiducial cosmology of~\citet{Planck_cosmo18} and predict the theoretical linear and non-linear power spectra in the \lcdm\ model. The non-linear power spectra uses \halofit~\citep{halofit1,halofit2} to model the non-linear matter power spectrum and we keep using a linear galaxy bias. We then determine the transition between the linear and non-linear regimes to correspond to the largest multipole $\ell_\textrm{max}$ such that the relative difference between the linear and non-linear power spectra is inferior to $5\%$. The determined scales are presented in Tab.~\ref{tab:scale_Cuts}. In practice we use $\ell_\textrm{max}=200$ for CMASS,  $\ell_\textrm{max}=100$ for LOWZ, and $\ell_\textrm{max}=400$ for CMB lensing. 

\begin{table}
    \center
    \begin{tabular}{cc}
    \hline
    Angular power spectrum & $\ell_\textrm{max}$ \\ 
    \hline
    \\[-0.95ex]
    $C_\ell^{gg,\textrm{CMASS}}$ & $202$                \\
    $C_\ell^{\kappa g,\textrm{CMASS}}$ & $208$ \\
    $C_\ell^{gg,\textrm{LOWZ}}$ & $106$ \\
    $C_\ell^{\kappa g,\textrm{LOWZ}}$ & $108$ \\
    $C_\ell^{\kappa \kappa}$ & $418$ \\
    \\[-0.95ex]
    \hline
    \end{tabular}
    \caption{Scale cuts for each power spectrum such that the relative difference between the linear and non-linear power spectra is inferior to $5\%$.}
    \label{tab:scale_Cuts}
\end{table}

Finally, since we use the Limber approximation, which is not valid on large scales, we limit our study to multipoles above $\ell_\textrm{min}=20$.

\subsection{Covariance matrix}

We use a Gaussian covariance matrix following ~\citet{Saraf22}. This allows us to incorporate non-overlapping regions of the sky within our cross-correlation analysis,
\begin{multline}\label{eq:covariance}
    {\rm Cov}_{LL'}^{AB,CD} = \frac{\delta_{LL'}}{(2\ell_L+1)\Delta \ell f_\textrm{sky}^{AB}f_\textrm{sky}^{CD}}\biggl[ f_\textrm{sky}^{AC,BD}\left(C_L^{AC}+N_L^{AC}\right) \\
    \times\left(C_L^{BD}+N_L^{BD}\right)+ f_\textrm{sky}^{AD,BC}\left(C_L^{AD}+N_L^{AD}\right)\left(C_L^{BC}+N_L^{BC}\right)\biggr],
\end{multline}
where $A$,$B$,$C$,$D$ label one of the two galaxy density fields $g$ or the CMB lensing convergence field $\kappa$, and $f_\textrm{sky}^{AB}$ is the  sky fraction common to fields $A$ and $B$. 

We estimate an initial covariance matrix using Eq.~(\ref{eq:covariance}) with the observed power spectra. This covariance matrix is used to fit our theoretical model to the observations, and we then determine a second covariance matrix using the best fit theoretical power spectra from the first analysis. This procedure reduces our sensitivity to the noise in the estimated power spectra.

\subsection{Likelihood}

Our log-likelihood is the sum of the log-likelihood of the galaxy power spectra and the galaxy -- CMB lensing cross-correlations, which we denote here as $\ln{\mathcal{L}^{2\times 2\,\textrm{pt}}}$, the \Planck\ log-likelihood for temperature and polarization~\citep{Planck_likelihood} $\ln{\mathcal{L}^{\textrm{Planck}}}$, and the \Planck\ CMB lensing~\citet{Aghanim20} likelihood $\ln{\mathcal{L}^{\textrm{Planck lensing}}}$,
\begin{align}\label{eq:likelihood}
    \ln{\mathcal{L}} = \ln{\mathcal{L}^{2\times 2\,\textrm{pt}}} + \ln{\mathcal{L}^{\textrm{Planck}}} + \ln{\mathcal{L}^{\textrm{Planck lensing}}}.
\end{align}

The name \twotimestwo\ comes from the fact that it is the combination of two different types of two point statistics ($C_\ell^{gg}$ and $C_\ell^{\kappa g}$). The third two point statistic is the \Planck\ lensing power spectrum, $C_\ell^{\kappa \kappa}$, making our analysis a \threetimestwo\ analysis in combination with the Planck temperature and polarisation likelihood. 

We adopt a Gaussian for the \twotimestwo\ likelihood,
\begin{align}
\ln{\mathcal{L}^{2\times 2\,\textrm{pt}}} = -\frac{1}{2}\left[\left(X(\theta)-X^\textrm{obs}\right)^TC^{-1}\left(X(\theta)-X^\textrm{obs}\right)\right],
\end{align}
where $X^T$ denotes the transpose of vector $X$, $\theta$ is the parameter vector, $C$ is the covariance matrix, and $X$ is a concatenation of power spectra such that,
\begin{align}
    X = \left(C_\ell^{gg,\textrm{CMASS}},C_\ell^{\kappa g,\textrm{CMASS}},C_\ell^{gg,\textrm{LOWZ}},C_\ell^{\kappa g,\textrm{LOWZ}}\right).
\end{align}

The likelihood $\mathcal{L}^{\textrm{Planck}}$ contains the likelihood for the TT, TE and EE power spectra. We have neglected the covariance between $C_\ell^{\kappa\kappa}$ and $C_\ell^{\kappa g}$ ~\citep[as is done, for instance, in][]{Abbott23}, which is motivated by the fact that the $C_\ell^{\kappa\kappa}$ is estimated on a much larger sky fraction than $C_\ell^{\kappa g}$, and that CMB lensing power comes mainly from redshifts greater than $0.8$, where we do not have any galaxy in either of the two samples.

\subsection{Priors}
We use flat priors for all of the cosmological parameters, namely $\omega_m$, $\omega_b$, $h$, $\tau_\textrm{reio}$, $\log{10^{10}A_s}$, $n_s$, as well as for the two galaxy bias parameters, for LOWZ and CMASS, and $\logfr$, for which we impose $-7<\logfr<0$. We use the recommended priors for the calibration and nuisance parameters required by \Planck's likelihood. In total, we end up with $30$ free parameters (the $6$ parameters of \lcdm, $\logfr$, $2$ galaxy bias parameters and $21$ calibration and nuisance parameters). We then use the MCMC (Monte Carlo Markov Chain) sampler \emcee\footnote{https://emcee.readthedocs.io/}~\citep{emcee} to sample the resulting posterior distributions.

\subsection{Computing prior-independent constraints}\label{sec:methodology_prior_effect}
As will be seen in Sect.~\ref{sec:results}, in many cases we are only able to put upper bounds on $\logfr$. It is then non-trivial to estimate the $95\%$ constraint on $\logfr$, since the percentile depends strongly on the lower value of the prior. This arises because the posterior on $\logfr$ at low values is non zero (and almost flat). Therefore, without a lower bound on the prior, very low values of $\logfr$ would be explored, rather than more interesting regions of the posterior, which would in turn be poorly sampled. This choice of  lower limit for $\logfr$ changes the estimated $95^\textrm{th}$ percentile. 

In order to resolve this issue, we follow the approach of~\citet{Piga23}, who rely on~\citet{Gordon07}. We consider the ratio of the marginalized posterior and our prior,
\begin{align}
    b(x; d, p) = \frac{\mathcal{P}(x\lvert d, p)}{p(x)},
\end{align}
where $x$ is the parameter we wish to constrain (in our case $\logfr$), $d$ is the data, $p$ the prior, and $\mathcal{P}$ the posterior. Then, for two different values of $x$, say $x_1$ and $x_2$, we apply Bayes theorem to obtain,
\begin{align}
    \frac{b(x_1;d,p)}{b(x_2;d,p)} = \frac{\mathcal{L}(d\lvert x_1)}{\mathcal{L}(d\lvert x_2)}=B(x_1,x_2),
\end{align}
where $B$ is the Bayes factor and $\mathcal{L}(d\lvert x)$ is the marginalized likelihood of the data for $x$. The Bayes factor, which quantifies the support for the model with $x=x_1$ over the model with $x=x_2$, is therefore prior independent. 

\citet{Gordon07} showed that $B(x_1,x_2)=2.5$ means that the model with $x=x_1$ is favored compared to the model with $x=x_2$ at $95\%$. We then fix $x_1$ to its lower bound of $\logfr=-7$ and find the value of $x_2$ such that $B(x_1,x_2)=2.5$. In this way, we are able to compute $95\%$ confidence intervals which do not depend on the prior. We check this is indeed the case in Sect.~\ref{sec:results_prior_effect}, where we vary the prior used in the analysis.

\section{Results}\label{sec:results}

\begin{figure}
  \resizebox{\hsize}{!}{\includegraphics{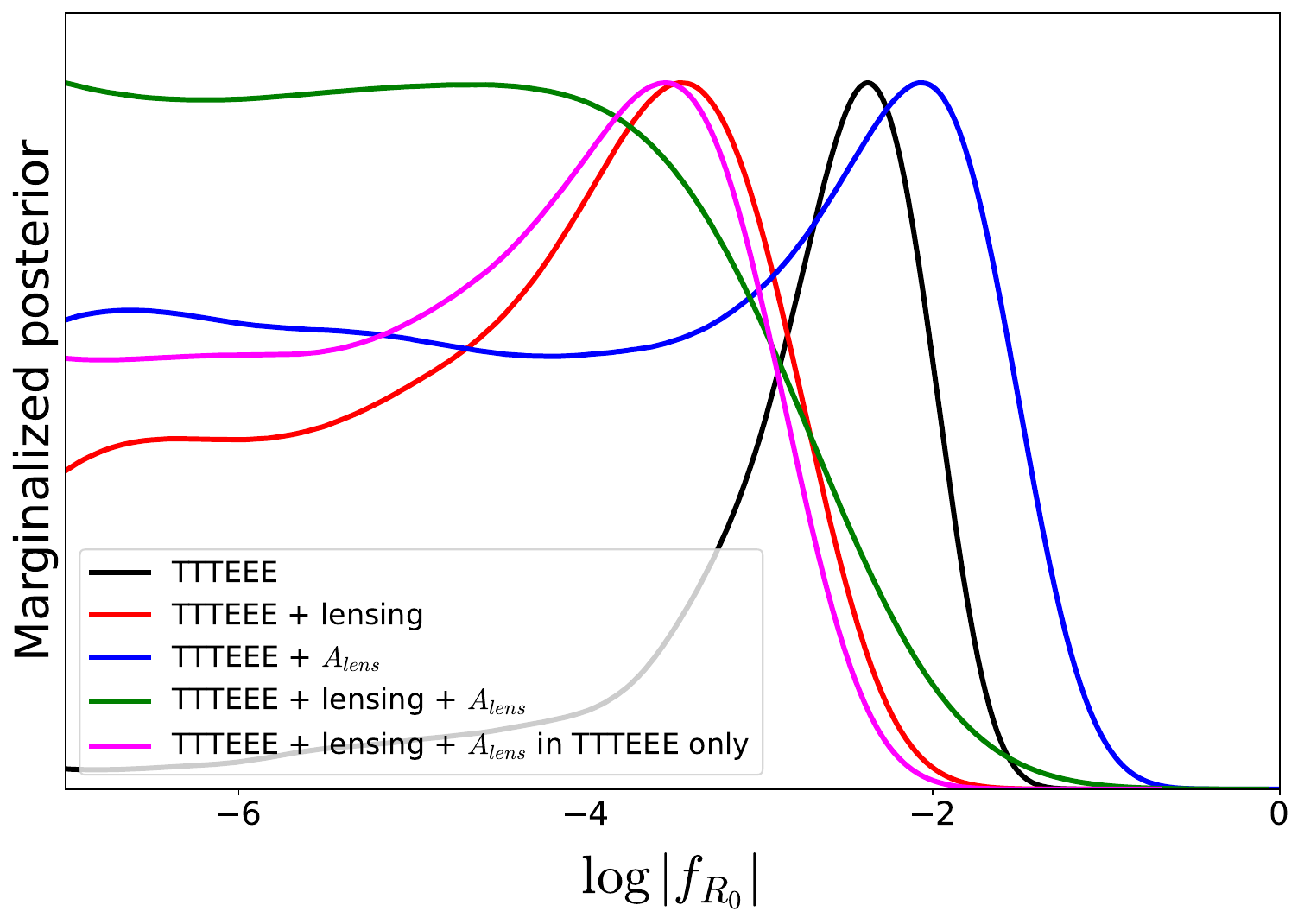}}
  \caption{Marginalized posterior of $\logfr$ using observations of \Planck\ modeled with \mgclass. TTTEEE refers to the temperature and polarization power spectra of \Planck, ‘‘lensing’’ to the addition of the CMB lensing power spectrum, and $\Alens$ to the inclusion of the lensing amplitude (see the text for  details).}\label{fig:constraint_fr0_Planck_Alens}
\end{figure}

In this section we present our results for different combinations of observations, with different choices of binnings, theoretical power spectra and nuisance parameters. The upper limits we place on $\logfr$ are shown in Tab.~\ref{tab:constraints_fr0} and, the corner plot for the fiducial \threetimestwo\ + CMB analysis, obtained thanks to the use of \textit{GetDist}\footnote{https://getdist.readthedocs.io/} \citep{2019arXiv191013970L}, is shown in Fig.~\ref{fig:triangle_plot_fr0_3x2}.

\subsection{Results from CMB only}\label{sec:results_CMB}

We first look at the constraints obtained on $\logfr$ when using only observations of the CMB. All these constraints are obtained using \mgclass, as \react\ can only make predictions for the matter power spectrum at redshifts $z<2.5$. 

Firstly, we use the CMB temperature and polarization power spectra. They are sensitive to \for\ gravity through the imprint of the  integrated Sachs–Wolfe effect and gravitational lensing. Lensing is the dominant effect and causes a smoothing of the acoustic peaks ~\citep[for more detail, see][]{Lewis06}. This is distinct from the CMB lensing convergence reconstruction described in Sect.~\ref{sec:lensing}. Since \for\ increases the power in the lensing potential, \for\ models will have a larger smoothing effect on the CMB.

With just the CMB temperature and polarization observations, we find a strong preference for a non-zero value of $f_{R_0}$ at more than $3\sigma$: a best-fit value of $\logfr = -2.35$, with the bounds $-3.09<\logfr<-1.88 $ at $95\%$ and $-5.76<\logfr<-1.64$ at $99.7\%$ confidence levels. The marginalized posterior for $\logfr$ is shown in Fig.~\ref{fig:constraint_fr0_Planck_Alens}, where we see a clear peak at $\logfr = -2.35$. This is not a new result and has been shown by other authors, for example~\citet{Dossett14} and~\citet{Hojjati16}.

However, when we add the CMB lensing power spectrum, which is estimated from the mode mixing in the primary CMB (as described in Sect.~\ref{sec:lensing}), the large values of $\logfr$ are excluded, and instead we set an upper limit on $\logfr$: $\logfr<-2.31$ at $95\%$ confidence (computed following the approach described in Sect.~\ref{sec:methodology_prior_effect}).

\begin{figure}
  \resizebox{\hsize}{!}{\includegraphics{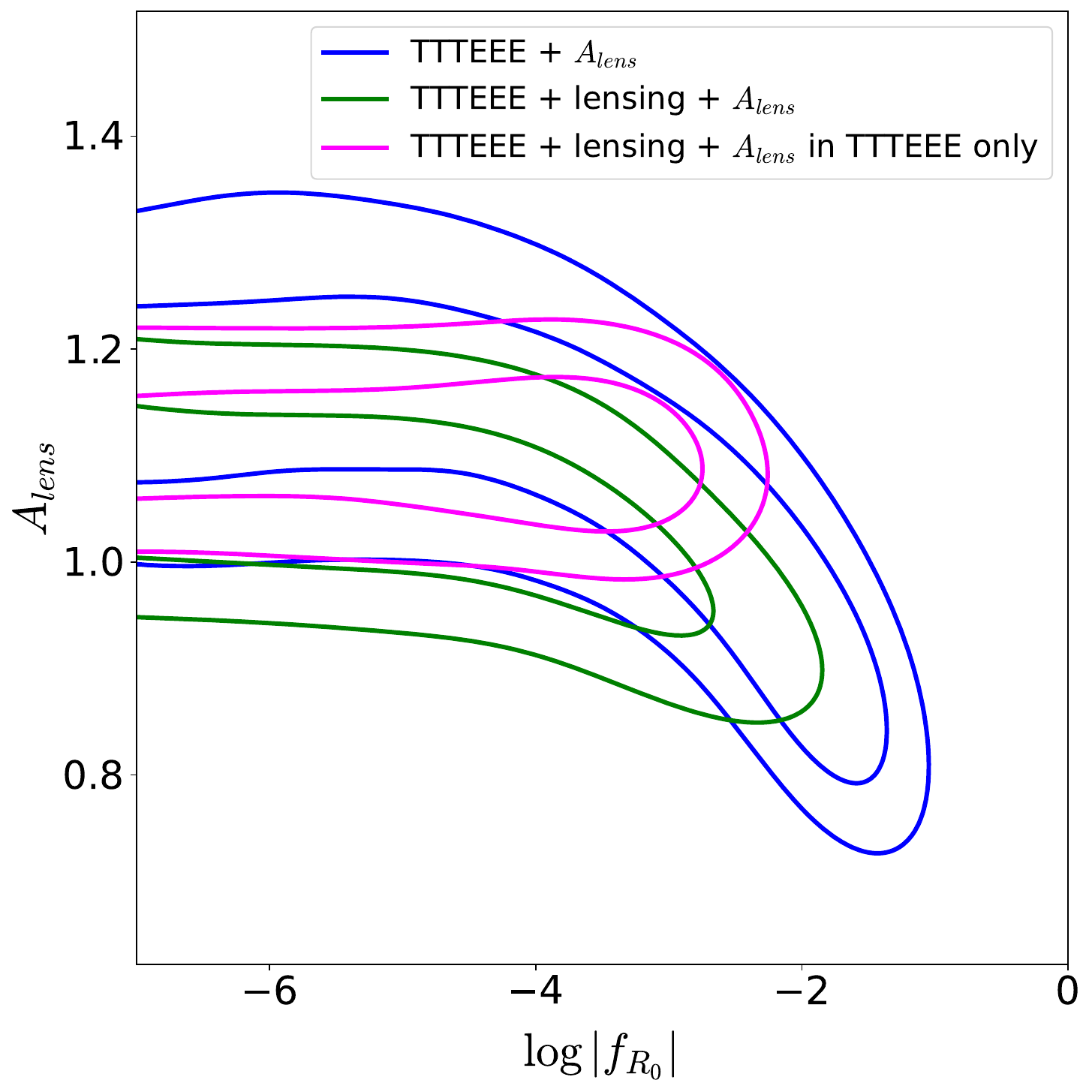}}
  \caption{Degeneracy between $\Alens$ and $\logfr$ for CMB temperature and polarization only (blue); CMB temperature, polarization and lensing (green); and when $\Alens$ is considered as a systematic effect and is applied only to the primary CMB anisotropies (magenta). The contours correspond to the $68^\textrm{th}$ and $95^\textrm{th}$ percentiles of the posterior samples.}\label{fig:Alens_vs_fr0}
\end{figure}

We clearly see that the CMB lensing convergence power spectrum is consistent with GR and low $\logfr$ values, while the smoothing of the acoustic peaks in the CMB temperature and polarization anisotropies prefers a higher value of $\logfr$. This issue (or tension) is closely related to the Planck $\Alens$ tension~\citep{Planck_cosmo18}.

The $\Alens$ parameter was introduced in~\citet{Calabrese08} as a phenomenological parameter scaling the CMB lensing potential amplitude as,
\begin{align}
    C_\ell^{\phi\phi}\rightarrow \Alens C_\ell^{\phi\phi}.
\end{align}
It therefore changes the amplitude of the CMB lensing convergence power spectrum, and also the smoothing in the temperature and polarization power spectra. It was shown in~\citet{Planck_cosmo18} that the CMB lensing convergence power spectrum is perfectly compatible with $\Alens=1$, which is not the case of the temperature and polarization power spectra. For instance, the $1\sigma$ constraint that~\citet{Planck_cosmo18} find using the TT,TE,EE+low E likelihood is $\Alens = 1.180 \pm 0.065$, which is in tension with $\Alens=1$ at $2.8\sigma$. Interestingly, the $\logfr$ tension is larger than the $\Alens$ tension. This suggests that the modification of the lensing potential by HS \for\ better describes the smoothing of the CMB than the simple rescaling of the potential with $\Alens$.

We further elucidate the relation between $\Alens$ and $\logfr$ by running the Planck CMB TTTEEE likelihood with both parameters.  Figure~\ref{fig:constraint_fr0_Planck_Alens} shows the result as the blue curve. The constraints on $\logfr$ are broadened, with low values of $\logfr$ now acceptable, and compatible with GR. The posterior, however, remains consistent with higher values of $\logfr$. Figure~\ref{fig:Alens_vs_fr0}  illustrates the degeneracy between $\Alens$ and $\logfr$.

The green and magenta curves in Fig.~\ref{fig:constraint_fr0_Planck_Alens} show the posteriors obtained when adding the CMB lensing convergence power spectrum likelihood. The difference between the two curves is that for the magenta curve, the effect of $\Alens$ is only  applied to the temperature and polarization power spectra; in this case, we are modeling $\Alens>1$ not as a physical effect, but rather as an unknown systematic in the observations. In both cases, the posteriors agree with GR, and we can put an upper limit on the value of $\logfr$.

We show the degeneracy between $\logfr$ and $\Alens$ in these two cases in Fig.~\ref{fig:Alens_vs_fr0}. When $\Alens$ is treated as a systematic (magenta contours), we recover high $\Alens$ values because the convergence power spectrum constrains $\logfr$ to low values where it cannot reproduce the smoothing observed in the acoustic peaks. These results clearly show that HS \for\ gravity cannot explain the $\Alens$ tension, in agreement with the findings of ~\citet{Hojjati16}.

\subsection{Results from combined CMB and \threetimestwo\ observations}\label{sec:results_fiducial}

\begin{figure}
  \resizebox{\hsize}{!}{\includegraphics{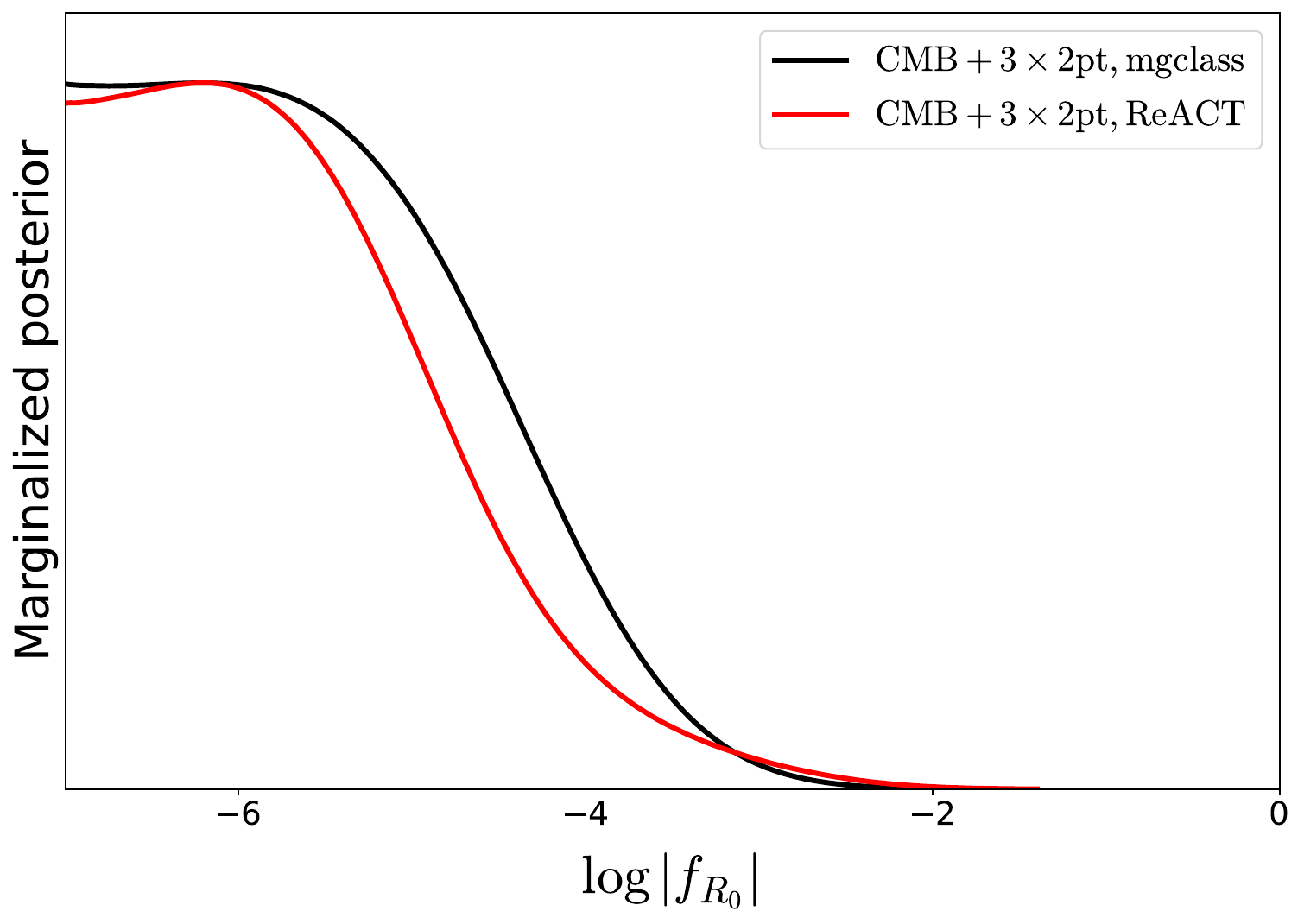}}
  \caption{Marginalized posterior of $\logfr$ using observations of CMB temperature and polarization, and the \threetimestwo\ observables (CMB lensing, galaxy distribution and their cross-correlation). The CMB observations including the CMB lensing auto power spectrum are always modeled using \mgclass, while the galaxy power spectra and the galaxy -- CMB lensing cross-correlations are modeled with \react\ (red) or \mgclass\ (black). }\label{fig:constraint_fr0_3x2}
\end{figure}

\begin{figure*}
   \centering
   \includegraphics[width=17cm]{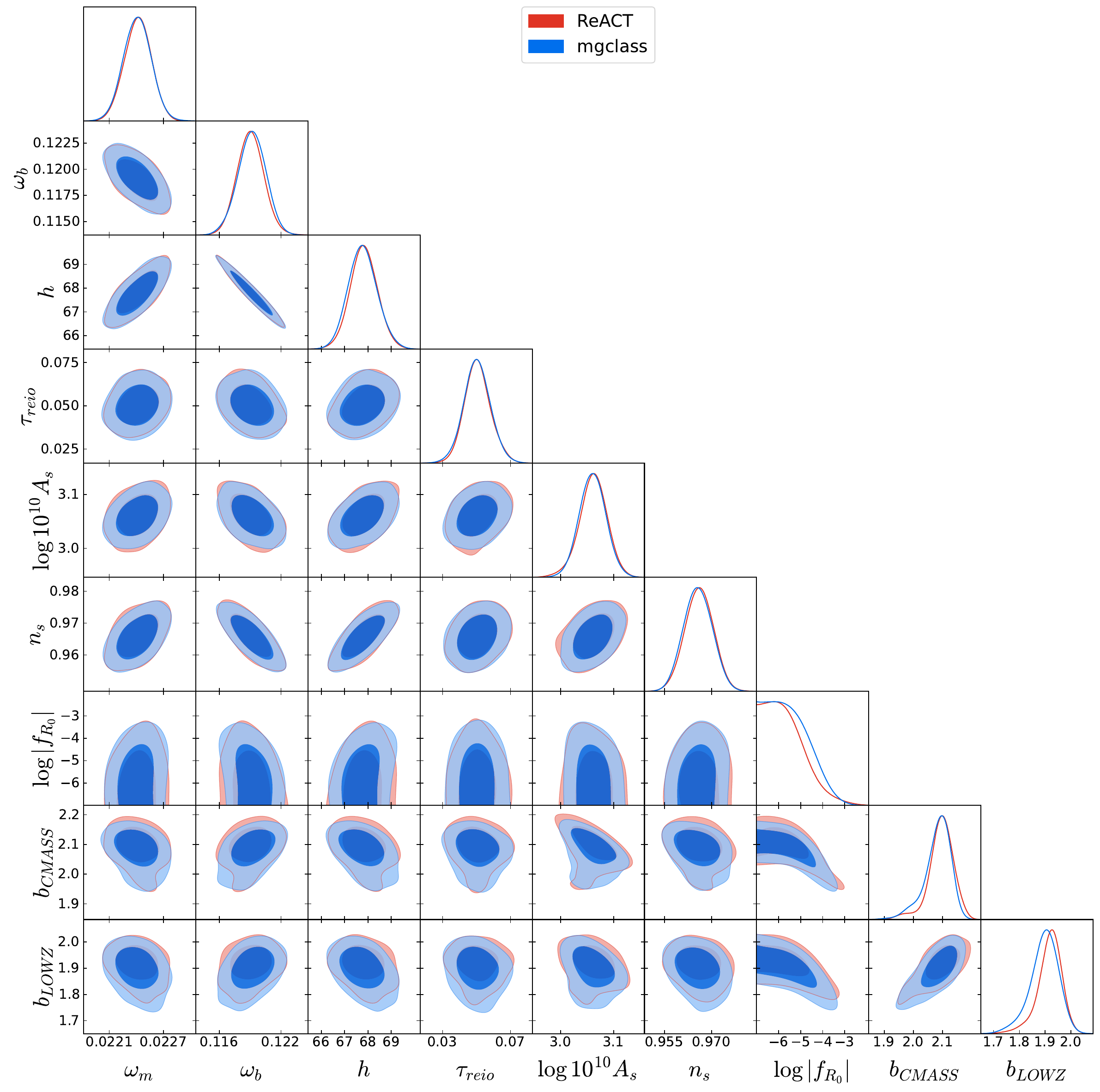}
   \caption{Constraints and degeneracies on all parameters when using observations of the CMB and \threetimestwo\ observables, modeled with \mgclass\ (blue) or \react\ (red). The contours correspond to the $68^\textrm{th}$ and $95^\textrm{th}$ percentiles of the posterior samples.}\label{fig:triangle_plot_fr0_3x2}
\end{figure*}

We now present our fiducial analysis, the combination of the CMB (temperature and polarization power spectra) and the \threetimestwo\ analysis (CMB convergence power spectrum, galaxy power spectrum and the cross-correlation between the two). In this section, the value of $\Alens$ is fixed to $1$ if not stated otherwise. 

Figure~\ref{fig:constraint_fr0_3x2} gives the marginalized posterior of $\logfr$ obtained when using either \mgclass\ (black curve) or \react\ (red curve) when modeling the \twotimestwo\ observables ($C_\ell^{gg}$ and $C_\ell^{\kappa g}$) and we recall that the CMB lensing convergence auto power spectrum is always modeled using \mgclass. The $95\%$ confidence level constraints are $\logfr<-4.12$ and $\logfr<-4.61$ with \mgclass\ and \react, respectively. These constraints are  consistent with GR and are much tighter than when using only the CMB observations. This also excludes HS \for\ as an explanation of the $\Alens$ tension. 

The difference between \mgclass\ and \react\ is also evident in Fig.~\ref{fig:cl_gg_variations_fr}. \mgclass\ only predicts the linear matter power spectrum, whereas \react\ predicts the non-linear power spectrum. Although our scale cuts were chosen to minimise the effects of non-linear structure formation, it seems likely that the origin of this difference here arises from the mildly non-linear regime where the \mgclass\ predictions have less power than those of \react. Therefore, our \mgclass\ constraint is likely too conservative and hence not as strong as it should be.

Figure~\ref{fig:triangle_plot_fr0_3x2} shows the marginalized two-dimensional posteriors on all parameters when using \mgclass\ (blue) or \react\ (red). We see that the posteriors agree well for most parameters. We also see that $\logfr$ exhibits a degeneracy with the galaxy bias parameters. This arises because both parameters change the amplitude of the galaxy power spectra. The parameters are not completely degenerate, however, as $\logfr$ also changes the shape of the power spectra, and the cross-correlation separates the two effects to a certain extent, as explored in the following subsection.

\begin{figure*}
   \centering
   \includegraphics[width=17cm]{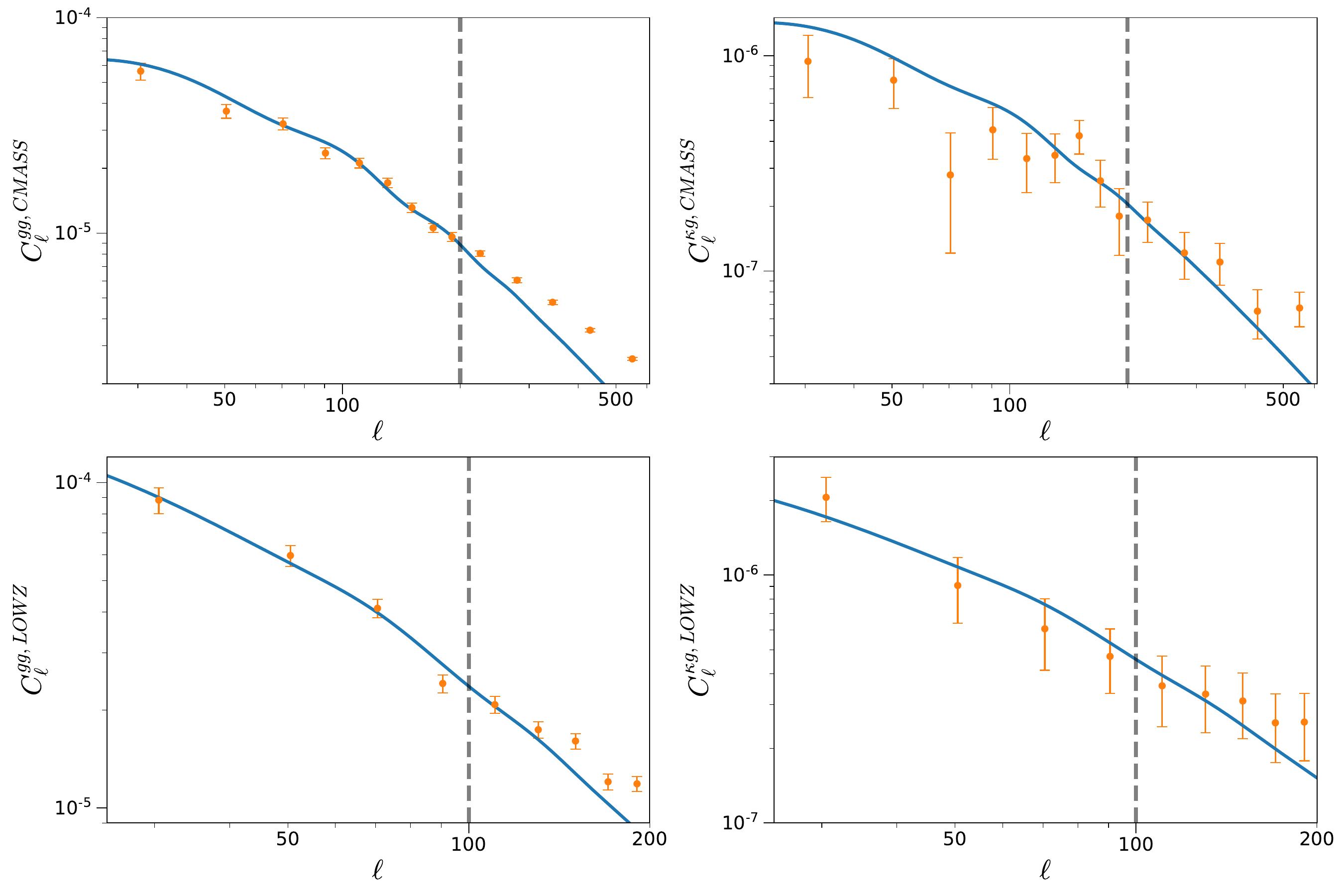}
  \caption{Measured angular power spectra (orange) and best fit (blue) obtained using \mgclass, as a function of multipole. The grey dashed line represents the limit between the linear and non-linear regimes. We only used multipoles above this limit in this analysis. Note the different multipole ranges for CMASS and LOWZ.}\label{fig:best_fits_mgclass}
\end{figure*}

We show in Fig.~\ref{fig:best_fits_mgclass} the measured angular power spectra (orange) and the theoretical best fit (blue) using \mgclass. The model fits well the data except at the largest scales for the cross-correlation between the CMB lensing of \Planck\ and the galaxies of CMASS, where the amplitude of the theoretical power spectrum is larger than the amplitude of the observation. This result has been observed in previous studies~\citep{Pullen16,Singh17,Kou22}. We see that HS \for\ is unable to resolve this problem.

\begin{figure}
  \resizebox{\hsize}{!}{\includegraphics{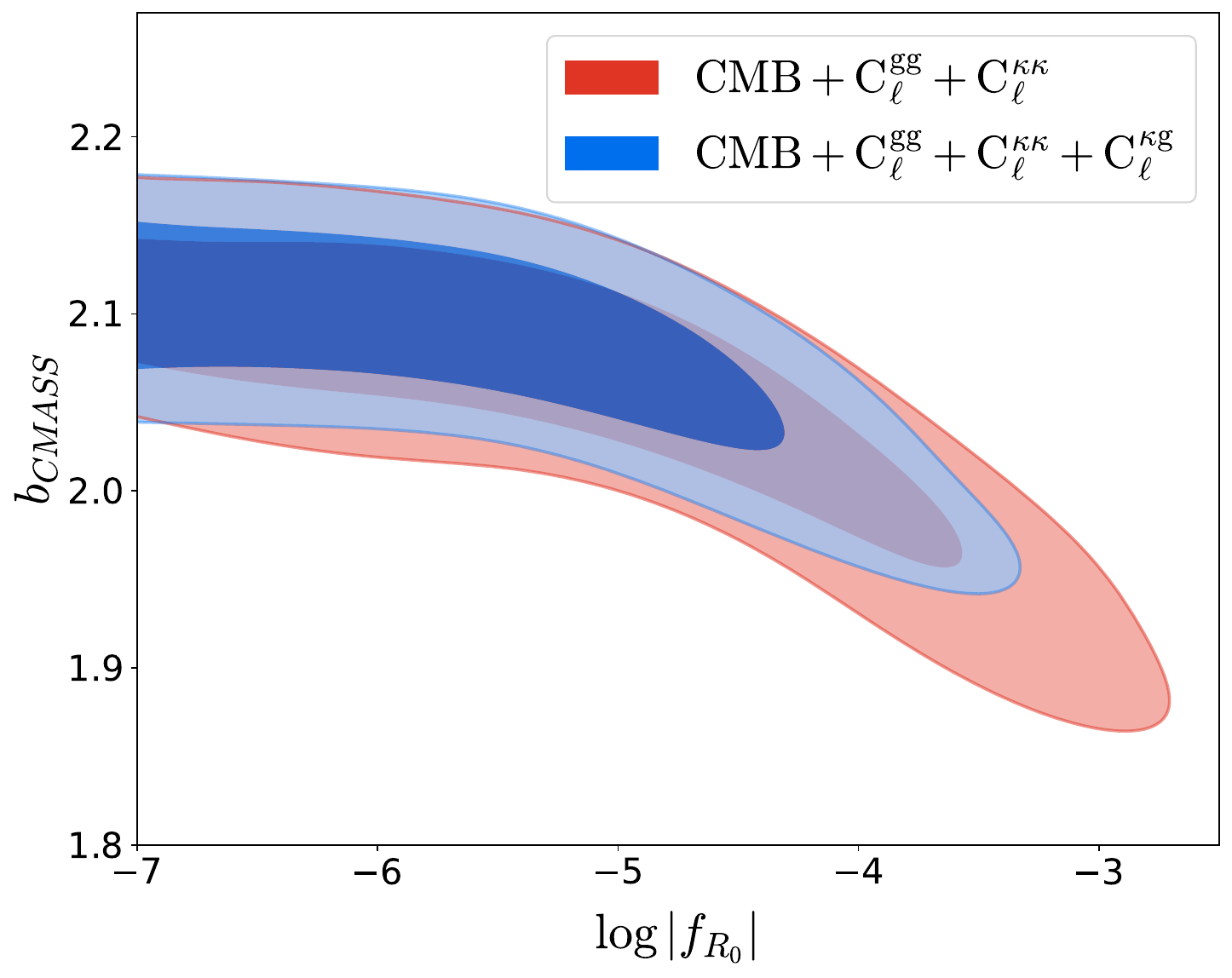}}
  \caption{Marginalized constraints on $\logfr$ and the galaxy bias of CMASS, when using CMB observations and the full \threetimestwo\ observables (blue) or only the galaxy and CMB lensing auto power spectra (red). It can be seen that adding the galaxy -- CMB lensing cross-correlation reduces the degeneracy between the two parameters. Those contours correspond to the $68^\textrm{th}$ and $95^\textrm{th}$ percentiles of the posterior samples.}\label{fig:constraint_fr0_bias}
\end{figure}

The constraints on $\logfr$ are consistent and competitive with previous studies using galaxy clustering observations, such as~\citet{Hu16}. Their tightest constraint is $\logfr<-4.5$ when combining observations of the CMB (temperature, polarization, lensing), supernovae, baryon acoustic oscillation (BAO) measurements (including, but not limited to, BAO measurements of LOWZ and CMASS) and galaxy weak lensing shear correlation functions estimated from the Canada-France-Hawaii Telescope Lensing Survey~\citep{Heymans13}. The cross-correlation of galaxy -- CMB lensing spectra enables us to obtain competitive constraints with a reduced data-set.

\subsubsection{Benefit of the cross-correlation}\label{sec:results_cross_correlation}

In order to isolate the advantage of the cross-correlation, we performed the analysis without the cross-correlation power spectra. The cross-correlation is primarily useful in reducing the degeneracy between $\logfr$ and the galaxy bias parameters. The degeneracy is reduced as $C_\ell^{gg}$ is proportional to $b_g^2$ while $C_\ell^{\kappa g}$ is proportional to $b_g$ (see Eq.~\ref{eq:cl_gg} and Eq.~\ref{eq:cl_kg}). This effect is seen in Fig.~\ref{fig:constraint_fr0_bias} where the red contours do not contain the cross-correlation and the blue contours do. As a result, the $95\%$ constraint obtained without the cross-correlation is $\logfr<-2.95$, compared to $\logfr<-4.12$ with the cross-correlation, i.e., more than an order of magnitude improvement.

\subsubsection{Result including $\Alens$ as a systematic}

\begin{figure}
  \resizebox{\hsize}{!}{\includegraphics{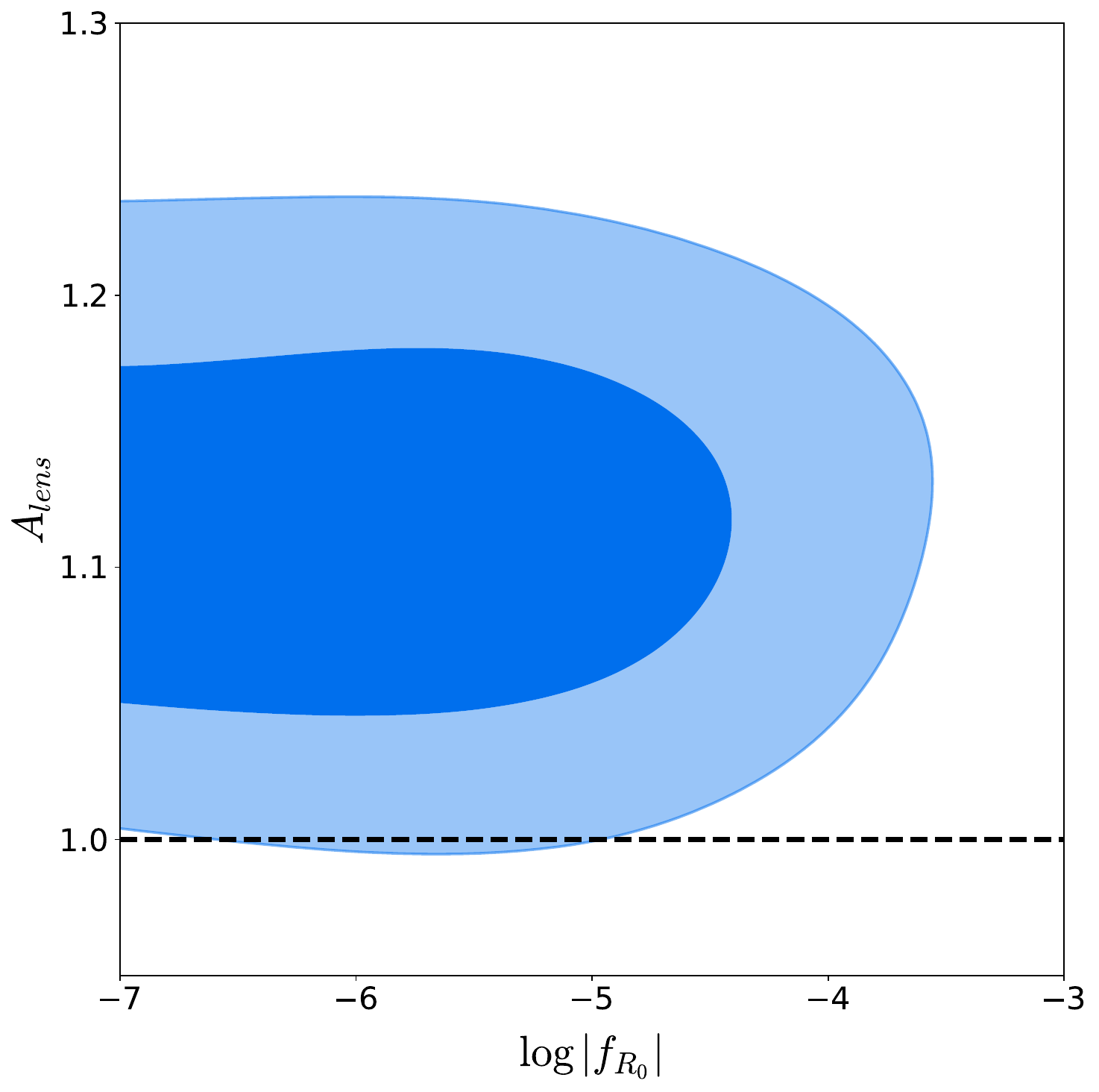}}
  \caption{Marginalized constraints on $\logfr$ and $\Alens$, where $\Alens$ is considered as a systematic effect and is only applied to the CMB temperature and polarization anisotropies. Here, the \threetimestwo\ observables are included, but are not affected by $\Alens$. A value of $\Alens>1$ is still preferred. The contours correspond to the $68^\textrm{th}$ and $95^\textrm{th}$ percentiles of the posterior samples.}\label{fig:constraint_3x2_fr0_Alens}
\end{figure}

As was shown in Sect.~\ref{sec:results_CMB}, HS \for\ cannot explain the $\Alens$ tension. When the \twotimestwo\ observations are added, $\logfr$ is constrained to even smaller values, ruling out further this kind of modified gravity resolution to the $\Alens$ tension. Since it has been suggested that $\Alens$ could be due to a systematic error~\citep{Planck_cosmo18}, rather than a physical effect, we also perform our analysis with $\Alens$ on the CMB temperature and polarization power spectra only. We expect this to give tighter constraints, because $\Alens$ can explain the excess of smoothing in the CMB temperature and polarization power spectra without the need for a large value of $\logfr$. The resulting marginalized contours on $\Alens$ and $\logfr$ are shown in Fig.~\ref{fig:constraint_3x2_fr0_Alens}. As expected, $\Alens>1$ is preferred. Unsurprisingly, the $95\%$ constraint of $\logfr<-4.24$  is tighter than the fiducial constraint ($\logfr<-4.12$).

\subsubsection{Effect of the binning scheme}

\begin{figure}
  \resizebox{\hsize}{!}{\includegraphics{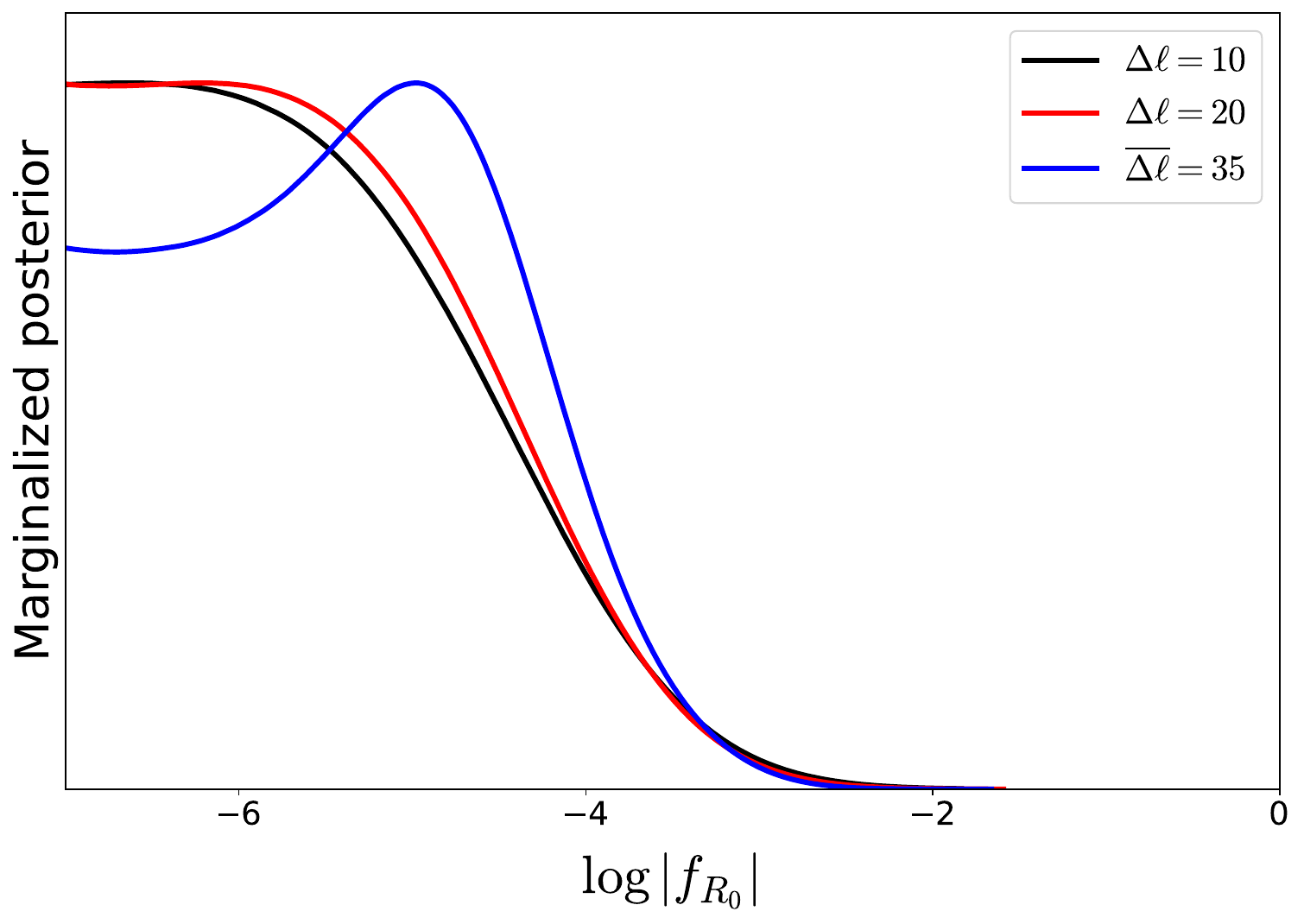}}
  \caption{Marginalized posterior of $\logfr$ using \mgclass\ depending on the binning scheme used. The constraints become tighter as $\Delta \ell$ decreases.}\label{fig:constraint_fr0_delta_l}
\end{figure}

We examine how changing the binning scheme impacts our constraints. In particular, we defined three binning schemes, namely $\Delta \ell = 10$, $\Delta \ell=20$ (the fiducial case) and a unequal binning scheme with average $\overline{\Delta \ell}=35$. The marginalized posteriors on $\logfr$ obtained with the three different binning schemes are presented in Fig.~\ref{fig:constraint_fr0_delta_l}. We see that the constraints become tighter as $\Delta \ell$ decreases. The use of narrow multipole bin-widths makes it possible to better use the shape of the power spectra to constrain \for\ gravity. The $95\%$ confidence constraints are $\logfr<-4.18$, $\logfr<-4.12$, and $\logfr<-3.84$ for $\Delta \ell = 10$, $\Delta \ell=20$, and $\overline{\Delta \ell}=35$, respectively. All of our constraints are summarized in Tab.~\ref{tab:constraints_fr0}.

\begin{table*}
    \center
    \begin{tabular}{cc}
    \hline
    Case &  Upper limit at $95\%$ confidence level  \\ 
    \hline
    CMB + $C_\ell^{\kappa\kappa}$ & $-2.31$ \\
    CMB + $C_\ell^{\kappa\kappa}$ + $\Alens$ & $-2.35$ \\
    CMB + $C_\ell^{\kappa\kappa}$ + $\Alens$ in CMB only & $-2.46$ \\
    Fiducial: CMB + $C_\ell^{\kappa\kappa}$ + $C_\ell^{gg}$ + $C_\ell^{\kappa g}$ (with \mgclass\ and $\Delta \ell = 20$)& $-4.12$ \\
    CMB + $C_\ell^{\kappa\kappa}$ + $C_\ell^{gg}$ + $C_\ell^{\kappa g}$ with \react\ & $-4.61$\\
    CMB + $C_\ell^{\kappa\kappa}$ + $C_\ell^{gg}$ + $C_\ell^{\kappa g}$, $\Delta \ell = 10$& $-4.18$ \\
    CMB + $C_\ell^{\kappa\kappa}$ + $C_\ell^{gg}$ + $C_\ell^{\kappa g}$, $\overline{\Delta \ell}=35$& $-3.84$ \\
    CMB + $C_\ell^{\kappa\kappa}$ + $C_\ell^{gg}$ & $-2.95$\\
    CMB + $C_\ell^{\kappa\kappa}$ + $C_\ell^{gg}$ + $C_\ell^{\kappa g}$ + $\Alens$ in CMB only & $-4.24$\\
    \hline
    \end{tabular}
    \caption{Constraints on $\logfr$ obtained in the different cases. CMB here refers to the temperature and polarization power spectra. The CMB alone, which is not included within this table, gives an incongruous constraint, reflecting the $\Alens$ tension, and is  discussed in the text in Sect.~\ref{sec:results_CMB}.}
    \label{tab:constraints_fr0}
\end{table*}

\subsubsection{Effect of the priors}\label{sec:results_prior_effect}

\begin{figure}
  \resizebox{\hsize}{!}{\includegraphics{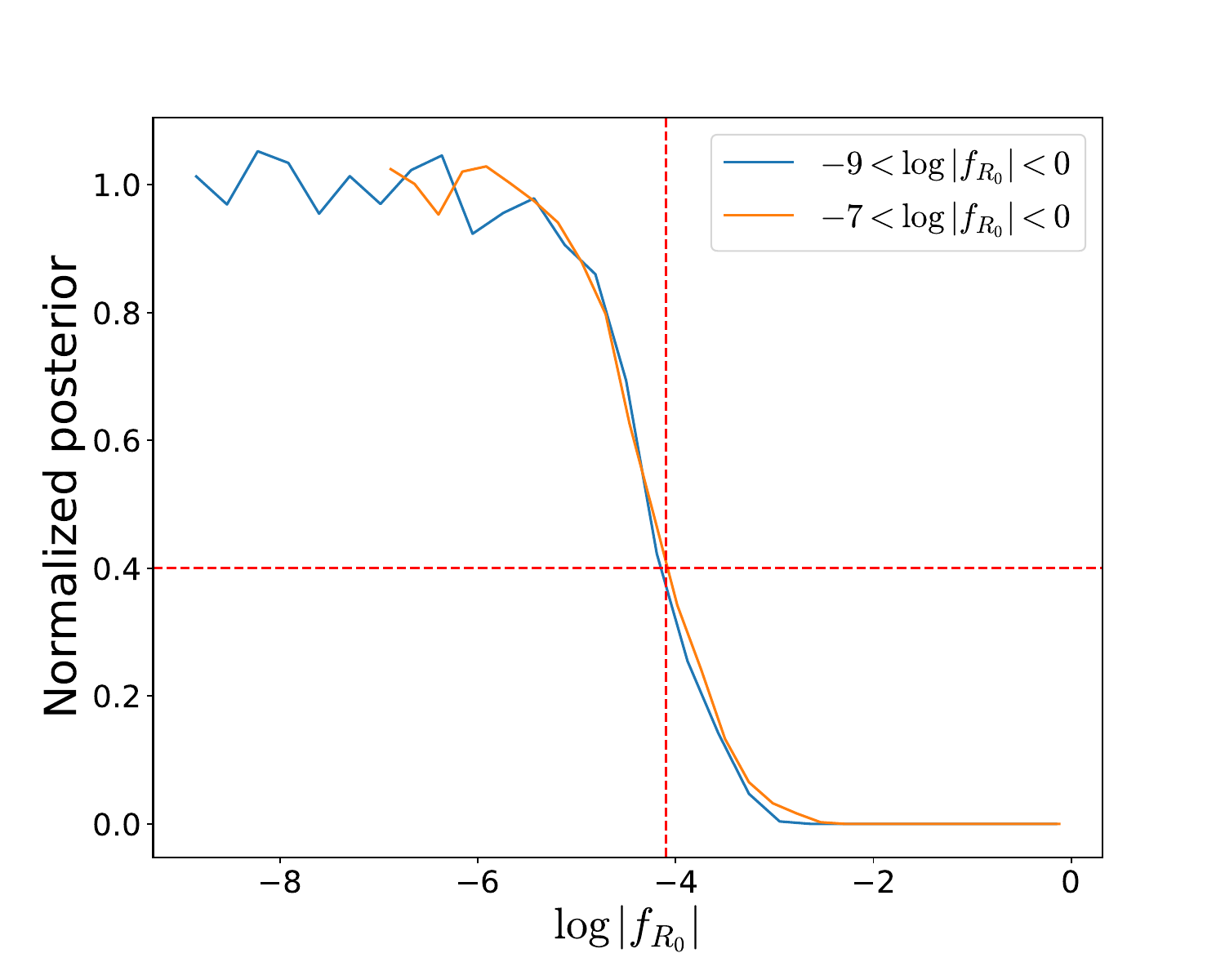}}
  \caption{Posteriors of $\logfr$ normalized by the average value of the plateau using flat priors between $-7$ and $0$ (orange) or between $-9$ and $0$ (blue). The red dashed lines correspond to the $95\%$ confidence level.}\label{fig:posterior_fr0_9_7}
\end{figure}

 Our $95\%$ confidence intervals rely on the approach described in Sect.~\ref{sec:methodology_prior_effect}, which aims at building prior-independent confidence intervals. In order to check that our constraints indeed are prior independent, we ran a second MCMC analysis in the fiducial setup, with a flat prior, $-9<\logfr<0$ (instead of the previous $-7<\logfr<0)$. The posteriors (normalized by the average value of the plateau) are shown in Fig.~\ref{fig:posterior_fr0_9_7}. It can be seen that the $95\%$ constraint is largely insensitive to the lower bound on the prior. More precisely, the constraint using the new prior is $\logfr<-4.17$, which is close to the limit $\logfr<-4.12$ that we found with the previous prior. 
 
 The approach described in Sect.~\ref{sec:methodology_prior_effect} indeed produces constraints independent of the adopted prior. In contrast, if we directly use the $95^\textrm{th}$ percentile of the samples, the values associated to each prior (lower bound at $-7$ or $-9$) would be, respectively, $-3.93$ and $-4.27$. This dependence on the choice of prior demonstrates the danger in determining the constraints directly from the sample percentiles; for instance, we extract a tighter constraint when the lower bound of the prior is lower. This finding is intuitive, since by decreasing the lower bound of the prior, we allow the MCMC walkers to explore lower values in parameter space; the posterior distribution thus shifts towards lower parameter values and, consequently, the $95^\textrm{th}$ percentile becomes lower as well.

\section{Conclusion} \label{sec:conclude}

Modified gravity is a possible explanation for the observed accelerated expansion of the Universe~\citep{Carroll04}. Hu-Sawicki (HS) \for\ gravity~\citep{Hu-Sawicki} is an attractive example, motivating the search for other possible observational signatures of the model. We searched for such signatures as deviations from the  predictions of General Relativity for large-scale structure observations in a combined  analysis of CMB, galaxy, and CMB lensing measurements.

If the HS \for\ model is to explain the accelerated expansion, the key parameter is $\logfr$. Primary CMB observations alone constrain this parameter through the ISW effect and smoothing of the temperature and polarization anisotropies by gravitational lensing. In agreement with previous analyses, we find that measurements by \Planck\ prefer high values of  $\logfr$, which would imply a remarkable deviation from General Relativity (see Sect.~\ref{sec:results_CMB}). However, this preference disappears when the CMB lensing convergence power spectrum is added, reflecting tension between the effects of lensing on the primary anisotropies and the reconstructed lensing power spectrum. This tension is closely related to the problem known as $\Alens$. We illustrate this by exhibiting the degeneracy between $\logfr$ and $\Alens$. This analysis also demonstrates that HS \for\ cannot by itself resolve this tension.

Setting $\Alens=1$ and adding galaxy power spectra from BOSS and their cross-correlation with CMB lensing, we then constrain $\logfr<-4.61$ at $95\%$ confidence (Sect.~\ref{sec:results_fiducial}). This is our central result. It means that while HS \for\ may still explain the accelerated expansion, there is no signature of the model in current observations of large-scale structure; the model predictions do not substantially deviate from those of General Relativity.  

We also showed that the cross-correlation of galaxy and lensing measurements is essential in breaking the degeneracy between galaxy bias and $\logfr$. Its addition improves the constraint on $\lvert f_{R_0} \lvert$ by  more than an order of magnitude (Sect.~\ref{sec:results_cross_correlation}). 

This paper is the first to make use of the cross-correlation between CMB lensing and galaxy measurements to constrain HS \for\ gravity. It paves the way for future large-scale galaxy surveys to place more stringent constraints, benefiting from lower noise and using galaxy lensing in addition to CMB lensing.

\begin{acknowledgements}
      We thank Benjamin Bose for his help in using the \react\ code. We acknowledge the use of the python libraries \textit{matplotlib} \citep{Hunter:2007}, \textit{numpy} \citep{harris2020array} and  \textit{scipy} \citep{2020SciPy-NMeth}. Some of the results in this paper have been derived using the {\sl healpy} and {\sl HEALPix} packages \citep{healpix}
\end{acknowledgements}

\bibliographystyle{aa} 
\bibliography{bibliography.bib}

\end{document}